\documentclass{PoS}

\def\bfOmega{\Omega \hspace{-9.1pt} \Omega}

\title{Effective Lagrangians on Domain Walls \\
and Other Solitons
}

\ShortTitle{Effective Lagrangians
}

\author{\speaker{Norisuke~Sakai}\thanks{Partly supported by 
Grant-in-Aid for Scientific Research from the Ministry of 
Education, Culture, Sports, Science and Technology, Japan 
No.17540237and No.18204024. 
}
\\
        Department of Physics, Tokyo Institute of 
Technology, 
Tokyo 152-8551, Japan 
\\
        E-mail: \email{sakai.n.aa(at)m.titech.ac.jp}}

\author{Minoru~Eto
\\
        University of Tokyo,  Inst. of Physics, 
Komaba 3-8-1, Meguro-ku 
Tokyo 153, Japan 
\\
        E-mail: \email{meto(at)hep1.c.u-tokyo.ac.jp}}

\author{Youichi~Isozumi
\\
        Department of Physics, Tokyo Institute of 
Technology, 
Tokyo 152-8551, Japan  
\\
        E-mail: \email{isozumi.y.aa(at)m.titech.ac.jp}}

\author{Muneto~Nitta
\\
        Department of Physics, Keio University, Hiyoshi,
Yokohama, Kanagawa 223-8521, Japan 
\\
        E-mail: \email{nitta(at)phys-h.keio.ac.jp}}

\author{Keisuke~Ohashi
\\
        Department of Physics, Tokyo Institute of 
Technology, 
Tokyo 152-8551, Japan  
\\
        E-mail: \email{keisuke(at)th.phys.titech.ac.jp}}

\abstract{Maintaining the preserved 
supersymmetry helps to find the effective Lagrangian on 
the BPS background in gauge theories with eight 
supercharges. 
As concrete examples, we take 1/2 BPS domain walls. 
The Lagrangian is given in terms of the superfields with 
manifest four preserved supercharges and is expanded in 
powers of the slow-movement parameter $\lambda$. 
The ${\cal O}(\lambda^0)$ gives the superfield form of the BPS 
equations, whereas all the fluctuation fields follow at 
${\cal O}(\lambda^1)$. 
The effective Lagrangian is given by the density of the 
K\"ahler potential which emerges automatically from the 
$\lambda$ expansion making four preserved supercharges 
manifest. 
More complete account of our method and applications 
is given in   \cite{Eto:2006uw} (hep-th/0602289) 
in which the case of non-Abelian vortices is also worked out.
}

\FullConference{From Strings to LHC\\
		January 2-10 2007\\
		International Centre Dona Paula, Goa India}

\begin{document}

\section{Introduction}\label{INTRO}
In the brane-world scenario, our 
four-dimensional spacetime is realized as a topological 
defect in a higher dimensional spactime 
\cite{HoravaWitten}.
It is desirable to construct such topological defect as 
various topological solitons. 
Supersymmetric gauge theories 
have been extremely useful to construct realistic models 
beyond the standard model \cite{DGSW}. 
When a field configuration preserves a part of 
supersymmetry (SUSY), it satisfies the field equation 
automatically \cite{WittenOlive}. 
Such a field configuration is called the 
Bogomol'nyi-Prasad-Sommerfield 
(BPS) state \cite{BPS}. 
The BPS solitons in the Higgs phase are extensively reviewed 
recently \cite{Eto:2006pg}. 
Quite often BPS solitons contain a number of parameters 
such as positions 
in the space-time and/or an internal space. 
These parameters are called moduli. 
To understand the dynamics of solitons for the brane-world 
scenario, it is important to construct the low-energy 
effective Lagrangian of the localized modes on such 
solitons. 
For that purpose, the standard method is to promote the 
moduli parameters of the background soliton into fields 
on the world volume of the soliton \cite{Manton:1981mp}. 
The moduli fields provide massless fields on the world 
volume of the soliton. 
This moduli approximation method is based on the assumption 
of the weak dependence on the world-volume coordinates, 
and gives the low-energy effective Lagrangian which 
contains all nonlinear terms with two derivatives or less. 
We have recently worked out a systematic method 
to obtain the effective Lagrangian on 
the BPS background in supersymmetric gauge theories, 
taking domain walls and vortices as concrete examples 
\cite{Eto:2006uw}. 
We have introduced a slow-movement parameter $\lambda$, 
and expanded the Lagrangian in terms of the superfields 
with four preserved supercharges in powers of the 
slow-movement parameter. 
We have found that maintaining the preserved supersymmetry 
manifest facilitates the procedure enormously. 
In this article, we introduce the systematic method taking 
domain walls as an example. 

We consider the supersymmetric $U(N_{\rm C})$ 
gauge theories with eight supercharges  
with $N_{\rm F} (\geq N_{\rm C})$ hypermultiplets in the 
fundamental representation as an illustrative example. 
Although we work in the space-time dimensions highest 
allowed by supersymmetry, namely domain walls in 
five dimensions, our discussion should be applicable 
in lower dimensions which can be obtained by dimensional 
reductions. 
We can naturally specify the order of magnitude in 
powers of the slow-movement parameter $\lambda$ for 
various fields. 
Thus we obtain a systematic expansion of the 
Lagrangian in powers of $\lambda$. 
The expansion gives 
a superfield form of the BPS equations at the 
zero-th order in $\lambda$, and the superfield equation to 
determine all the fluctuation fields at the next order. 
We retain up to the terms of order $\lambda^2$ 
in the Lagrangian, 
in order to obtain the effective Lagrangian 
at the lowest nontrivial order, namely up to two 
derivatives. 
We are now computing the higher powers of $\lambda$ 
in our systematic expansion to obtain the effective 
Lagrangian with higher derivative terms. 
We maintain four preserved SUSY manifest throughout 
our procedure, and obtain a density of the K\"ahler 
potential in four SUSY superspace. 
By integrating over the extra dimensions,
we obtain the K\"ahler potential of the effective Lagrangian 
which was difficult to obtain in general previously.
Our results can be used to study soliton scattering 
in $U(N_{\rm C})$ gauge theories.

\section{Slow-move Approximation in terms of Component Fields}
\label{sc:slow-move-comp}

Let us introduce our model and describe briefly the usual 
component method to solve the BPS equations. 
The bosonic parts of the Lagrangian with a 
common gauge coupling constant $g$ for 
$U(N_{\rm C}) = SU(N_{\rm C}) \times U(1)$  
in five dimensions is given by 
\begin{eqnarray}
{\cal L}|_{\rm boson} 
= 
{\rm Tr}\biggl[-{1\over 2g^2} F_{MN}(W)F^{MN}(W)
-\frac{1}{g^2}({\cal D}_M \Sigma)^2 
- {\cal D}^M H^i ({\cal D}_M H^i)^\dagger \biggr] - V
.
\label{fundamental-Lag2}
\end{eqnarray}
The physical bosonic components in the vector multiplet 
are gauge fields $W_M$, and the real scalar fields $\Sigma$ 
in the adjoint representation, 
and those in the hypermultiplet are the doublets of 
the complex scalar fields $H^i$ 
$i=1,2$ which can be assembled into 
$N_{\rm C}\times N_{\rm F}$ matrices. 
The indices $M, N=0, 1,\cdots, 4$ run over five-dimensions, 
and the mostly plus signature is used for the 
metric $\eta_{MN}={\rm diag}.(-1, +1, \cdots, +1)$. 
The covariant derivatives are defined as 
$D_M \Sigma = \partial_M \Sigma + i[ W_M , \Sigma ]$, 
$D_M H^{i}=(\partial_M + iW_M)H^{i}$, 
and field strength is defined as 
$F_{MN}=\frac{1}{i}[D_M , D_N]
=\partial_M W_N -\partial_N W_M + i[W_M, W_N]$. 
After eliminating auxiliary fields,  
the scalar potential $V$ is given by 
\begin{eqnarray}
\!\!\!
\!\!\!
\!\!\!
\!\!\!
V
=
\frac{g^2}{4}
{\rm Tr}
\Big[
\left(
H^{1}  H^{1\dagger}  - H^{2} H^{2\dagger} 
- c\mathbf{1}_{N_{\rm C}}
\right)^2 +
 4 H^2H^{1\dagger} H^1H^{2\dagger}
\Big] 
+{\rm Tr}\left[
 (\Sigma H^i - H^i M) 
 (\Sigma H^i - H^i M)^\dagger 
 \right]
\end{eqnarray}
with the hypermultiplet 
mass matrix $M={\rm diag}(m_1,\,\cdots, m_{N_{\rm F}})$ 
($m_A \in {\bf R}$)
and the Fayet-Iliopoulos parameter 
taken along the 
third direction in $SU(2)_R$ as 
$c_a =(0,\ 0,\ c)$ with $c>0$. 

By requiring half of SUSY to be preserved, we obtain 
the $1/2$ BPS equations for domain walls which depend on 
$y$ only 
\begin{eqnarray}
{\cal D}_y H^1 &=&
-\Sigma H^1 + H^1 M,\qquad 
{\cal D}_y H^2 = \Sigma H^2 -H^2 M,\label{BPSeq-H}
\\
{\cal D}_y \Sigma &=&
{g^2\over 2}\left(c{\bf 1}_{N_{\rm C}}-H^1H^1{}^\dagger 
+H^2H^2{}^\dagger \right), 
\label{BPSeq-Sigma}
\qquad 
0 = 
g^2 H^1H^2{}^\dagger . 
\end{eqnarray}
The solution of the BPS equations saturates the BPS bound for 
the tension of the (multi-)wall  
\begin{eqnarray}
T_{\rm w} &=&
\int^{+\infty}_{-\infty}\hspace{-1.5em}dy{\cal E}_{\rm w}
=\int^{+\infty}_{-\infty}\hspace{-1.5em}dy
\partial_y\Big[{\rm Tr}\big[c\Sigma 
-(\Sigma H^1H^{1\dagger }-H^1MH^{1\dagger })
+(\Sigma H^2H^{2\dagger }-H^2MH^{2\dagger })\big]\Big]
\nonumber \\
&=&
c \left[{\rm Tr}\Sigma \right]^{+\infty}_{-\infty}
\label{eq:tension}
\end{eqnarray}
where the energy density is denoted as ${\cal E}_{\rm w}$. 
We can solve the hypermultiplet BPS equation 
(\ref{BPSeq-H}) by introducing an $N_{\rm C}\times N_{\rm F}$ 
constant matrix $H_0$ called the {\it moduli matrix} 
\cite{Isozumi:2004jc}, \cite{Isozumi:2004vg}
\begin{eqnarray}
 \Sigma +iW_y&=&S^{-1}(y)\partial _yS(y), 
\qquad 
W_\mu =0, \qquad 
(\mu =0,\cdots,3) 
\label{eq:comp-gauge-tr} 
\\
H^1&=& S^{-1}(y)H_0e^{My}, \qquad H^2=0, 
\label{eq:hyper-sol} 
\end{eqnarray}
where the moduli matrix $H_0$ carries all the parameters 
of the solution, namely moduli. 
The moduli matrices related by the following $V$-equivalence 
transformations are physically equivalent: 
\begin{eqnarray}
 H_0 \rightarrow V H_0, 
\qquad 
S(y) \rightarrow V S(y), 
\qquad 
V \in GL(N_{\rm C}, {\bf C}). 
\label{eq:v_equivalence} 
\end{eqnarray}
The vector multiplet BPS equation (\ref{BPSeq-Sigma}) 
can be converted to the following 
``master equation'' for a gauge invariant 
quantity $\Omega \equiv SS^\dagger $ \cite{Isozumi:2004jc} 
\begin{eqnarray}
 \partial _y\left(\Omega ^{-1}\partial _y\Omega \right)
=g^2c\left({\bf 1}_{N_{\rm C}}-\Omega ^{-1}\Omega _0\right),
\qquad \Omega _0\equiv c^{-1}H_0e^{2My}H_0^\dagger. 
\label{eq:master-eq}
\end{eqnarray}
The matrix function $S$ can be determined from the solution 
$\Omega$ of this master equation by fixing a gauge, 
and all the other fields can be obtained from $S$ and $H_0$. 
Since the BPS soliton has co-dimension one, the solution 
represents (multiple parallel) domain walls. 

We can obtain the low-energy effective Lagrangian 
by promoting the moduli parameters $\phi ^\alpha $ in the 
moduli matrix $H_0$ to fields on the 
soliton world volume depending on $x^\mu , (\mu =0,1,2,3)$ 
\begin{eqnarray}
 H_0(\phi ^\alpha )\rightarrow H_0(\phi ^\alpha (x)) . 
\label{eq:moduli-fields}
\end{eqnarray}
To represent the weak dependence on the world-volume 
coordinates of the soliton \cite{Manton:1981mp}, 
we introduce ``the slow-movement parameter'' $\lambda$, 
which is assumed to be much smaller than the typical 
mass scale in the problem, in our case, the hypermultiplet 
mass difference $\Delta m$ and $g\sqrt{c}$ where 
$c$ and $g$ are the Fayet-Iliopoulos parameter and the gauge 
coupling.  
\begin{eqnarray}
 \lambda \ll {\rm min}(\Delta m, g\sqrt{c} ).
\end{eqnarray}
The nonvanishing fields of the $1/2$ BPS background have 
contributions independent of $\lambda$, and 
derivatives in terms of the world volume coordinates 
are assumed to be of order $\lambda$, expressing 
the weak dependence on the world-volume coordinates 
\begin{eqnarray}
H^1 \sim {\cal O}(1),\quad \Sigma \sim {\cal O}(1),\quad 
\partial _\mu \sim {\cal O}(\lambda ). 
\label{eq:h1-order}
\end{eqnarray}
Those fields which vanish in the background solution can 
now have nonvanishing values, induced by the fluctuations 
of the moduli fields of order $\lambda$ 
\begin{eqnarray}
&W_\mu \sim {\cal O}(\lambda ),\quad 
H^2\sim {\cal O}(\lambda ) ,& \nonumber \\
 &{\cal D}_\mu H^1\sim {\cal O}(\lambda ),
\quad {\cal D}_\mu \Sigma\sim {\cal O}(\lambda ),
\quad F_{\mu y}(W)\sim {\cal O}(\lambda ),& 
\end{eqnarray}
and other components of the field strength are higher 
orders in $\lambda$. 
If we decompose the field equations in powers of 
$\lambda$, we find that order $\lambda ^0$ equations 
are automatically satisfied by the BPS equations 
(\ref{BPSeq-H}) and (\ref{BPSeq-Sigma}). 
However, it becomes more and more complicated to 
solve the field equation at higher orders in the 
expansion in powers of $\lambda$, since various fields 
that vanish in the background become nonvanishing, and 
need to be solved. 
We shall show in the next section that maintaining the 
preserved SUSY manifest greatly helps to determine these 
newly nonvanishing fields and to organize the expansion of 
field equations in powers of $\lambda$.

\section{Slow-move Approximation in terms of Superfields 
}

Since four supercharges are conserved by the BPS domain 
walls,  an action for fluctuations around the BPS 
background can be written in term of the superfield for 
four supercharges. 
Let us define the superfields\footnote{ 
We use the convention of Wess and Bagger \cite{WessBagger} 
for Grassmann coordinates and superfields in this paper, 
except that four-dimensional 
spacetime indices are denoted by Greek alphabets 
$\mu, \nu=0, \cdots, 3$. 
For conventions of superfields in terms of component fields, 
we mostly follow those in Refs.\cite{Hebecker}, and 
\cite{KakimotoSakai}. } 
 using two component spinor 
Grassmann coordinates 
$\theta^\alpha, \bar\theta_{\dot \alpha}$. 
The components of superfields are fields in five dimensions. 
A vector multiplet with eight SUSY consists of 
a real vector superfield ${\bf V} (={\bf V}^\dagger)$ 
and an adjoint chiral superfield ${\Phi}$ 
($\bar D_{\dot\alpha} {\Phi} = 0$) 
in terms of superfield with four supercharges 
\cite{AGW}. 
The vector superfield ${\bf V}$ 
contains a gauge field $W_\mu, \mu=0, \cdots, 3$ 
for the four spacetime dimensions, the half of gaugino 
field $\lambda_+$, and an auxiliary field ${\cal Y}^3$. 
If one takes the Wess-Zumino gauge, it becomes explicitly as 
\begin{equation}
{\bf V}\Big|_{\rm WZ}=
-\theta \sigma^\mu \bar\theta W_\mu 
+ i\theta^2\bar \theta \bar \lambda_+
- i\bar \theta^2 \theta \lambda_+ 
+{1 \over 2} \theta^2\bar \theta^2 {\cal Y}^3, 
\qquad {\cal Y}^3\equiv Y^3-{\cal D}_y\Sigma,  
\label{eq:vector-superfield}
\end{equation}
where the auxiliary field ${\cal Y}^3$ of the superfield 
for four SUSY is shifted from the auxiliary field $Y^3$ 
for eight SUSY by the covariant derivative of adjoint 
scalar $\Sigma$ along the fifth coordinate (the extra 
dimensions) $y$ \cite{Hebecker}, \cite{KakimotoSakai}. 
This difference becomes important in identifying the 
topological charge later. 
The chiral scalar superfield ${\Phi}$ contains a 
complex scalar field made of the adjoint scalar $\Sigma$ 
and the fifth component of the gauge field $W_y$ as the 
real and imaginary parts respectively, and the other half 
of gaugino $\lambda_-$ and a complex auxiliary field 
$Y^1+iY^2$ 
\begin{equation}
{\Phi }
=\Sigma + i W_y + \sqrt{2}\theta (-i\sqrt{2} \lambda_-) 
+ \theta^2 (Y^1 +i Y^2). 
\label{eq:vectorr-chiral-superfield}
\end{equation}
The hypermultiplet are represented by a chiral superfield 
${\bf H}^1$ and an anti-chiral superfield ${\bf H}^2$. 
The (anti-) chiral superfield ${\bf H}^1$ (${\bf H}^2$) 
consists of the physical complex scalar field ${H}^1$ 
(${H}^2$), hyperino $\psi_+$ ($\psi_-$), and a complex 
auxiliary field ${\cal F}^1$ (${\cal F}^2$) 
\begin{eqnarray}
&& {\bf H}^1=H^1+\sqrt{2}\theta \psi_++\theta^2 {\cal F}^1, 
\qquad 
 {\cal F}^1\equiv F^1+({\cal D}_y-\Sigma )H^2+H^2M, 
 \label{eq:hyper-superfield1} \\
&&{\bf H}^2=H^2+\sqrt{2}\bar \theta \bar \psi_-
 +\bar \theta^2 {\cal F}^2, 
 \qquad 
 {\cal F}^2\equiv -F^2-({\cal D}_y+\Sigma )H^1+H^1M, 
 \label{eq:hyper-superfield2}
\end{eqnarray}
where the auxiliary field ${\cal F}^1$ (${\cal F}^2$) 
of the superfield for four SUSY is shifted from 
the auxiliary field $F^1$ ($F^2$) 
for eight SUSY by the covariant derivative of the 
other hypermultiplet scalar $H^2$ ($H^1$) and 
other\footnote{
The other terms involving the adjoint scalar $\Sigma$ 
and the hypermultiplet mass matrix $M$ can be 
understood as a result of the Scherk-Schwarz dimensional 
reduction from six dimensions. 
} 
terms~\cite{Hebecker}, \cite{KakimotoSakai}. 
Please note that we have chosen to denote the anti-chiral 
superfield as ${\bf H}^2$, as shown in the $\bar \theta$ 
dependence in Eq.(\ref{eq:hyper-superfield2}). 

The derivative $\hat D_y$ which is covariant 
under the complexified gauge transformations 
for the hypermultiplet 
${\bf H}^1$ and the adjoint chiral scalar multiplet 
${\Phi}$ are given by 
\begin{eqnarray}
{\hat D}_y{\bf H}^1
=(\partial _y+{\Phi}){\bf H}^1,  \quad\quad
{\hat D}_ye^{2{\bf V}} = 
\partial_y e^{2{\bf V}}-{\Phi}^\dagger e^{2{\bf V}}-
e^{2{\bf V}}{\Phi}. 
\end{eqnarray}
If supplemented by fermionic terms, the 
bosonic Lagrangian 
(\ref{fundamental-Lag2}) becomes invariant under the 
supersymmetric transformations with eight (real) 
Grassmann parameters. 
We can now rewrite this fundamental Lagrangian ${\cal L}$ 
in terms of the superfields for four supercharges as 
\begin{eqnarray}
 {\cal L}
&=&-{\cal E}_{\rm w} 
+ \int d ^4\theta{\rm Tr}\left[
-2c{\bf V}
+{1\over 2g^2}\left(
e^{-2{\bf V}}{\hat D}_ye^{2{\bf V}}\right)^2
+e^{2{\bf V}}{\bf H}^1{\bf H}^{1\dagger }
+e^{-2{\bf V}}{\bf H}^2{\bf H}^{2\dagger }
\right]\nonumber \\
&&{}+\left(\int d ^2\theta 
{\rm Tr}\left[
{\hat D}_y{\bf H}^1{\bf H}^{2\dagger }
-{\bf H}^1M{\bf H}^{2\dagger }
+{1\over 4g^2}{\bf W}^\alpha {\bf W}_\alpha 
\right]+{\rm h.c.}\right), 
\label{eq:fund-Lag-5d-4susy}
\end{eqnarray}
where field strength superfield ${\bf W}$ is given by 
\begin{eqnarray}
{\bf W}_\alpha \equiv -{1 \over 8} \bar D \bar 
D e^{-2{\bf V}} D_\alpha e^{2{\bf V}} . 
\end{eqnarray}
In transforming the fundamental Lagrangian 
(\ref{eq:fund-Lag-5d-4susy}) in terms of the 
superfield for four SUSY into the manifestly supersymmetric 
form for eight SUSY (\ref{fundamental-Lag2}), 
we need to make several partial 
integrations with respect to the fifth coordinate $y$, and 
have to retain the surface terms carefully 
in the procedure. 
We also note that the 
auxiliary fields for four SUSY ${\cal Y}^3$, 
and ${\cal F}^i$ are 
different from those for eight SUSY $Y^3$, $F^i$ 
by total derivative terms as in 
Eqs.(\ref{eq:vector-superfield}), 
(\ref{eq:hyper-superfield1}),  and 
(\ref{eq:hyper-superfield2}). 
In this way we find a total divergence ${\cal E}_{\rm w}$ 
representing the topological charge contributing to the 
energy density of the background which maintains four SUSY. 
Since we are interested in bosonic components of the 
topological term ${\cal E}$, we exhibit only the bosonic 
terms explicitly 
\begin{eqnarray}
{\cal E}_{\rm w}
&=&\partial_y\Big[{\rm Tr}\big[c\Sigma 
-(\Sigma H^1H^{1\dagger }-H^1MH^{1\dagger })
+(\Sigma H^2H^{2\dagger }-H^2MH^{2\dagger })\nonumber \\
&& {}
-{2\over g^2}{\cal Y}^3\Sigma 
+{\cal F}^1H^{2\dagger }+H^2{\cal F}^{1\dagger }
+({\rm fermionic~terms})\big]\Big]
.
\label{eq:total-der}
\end{eqnarray}
Let us emphasize again that the topological term is precisely 
the difference between the fundamental Lagrangian which 
is manifestly supersymmetric under the eight SUSY and 
another fundamental Lagrangian in terms of superfields 
for four manifest SUSY.

Following the usual procedure\cite{Manton:1981mp},  
we promote the moduli $\phi^i$ to fields $\phi^i(x)$ on 
the world volume of the background soliton, and 
assume that the moduli fields $\phi ^i(x)$ around the 
wall background to fluctuate only very slowly. 
Namely, we introduce a parameter $\lambda$ for the 
slow movement and neglect high energy fluctuations 
as explained in sect.\ref{sc:slow-move-comp}. 
By explicitly writing the derivatives of moduli fields 
we obtain 
\begin{eqnarray}
 \partial _y\phi ^i={\cal O}(1)\phi ^i,\qquad 
\partial _\mu \phi ^i={\cal O}(\lambda )\phi ^i,\quad  
\lambda \ll {\rm min}(\Delta m, g\sqrt{c}). 
\label{eq:slow-move1}
\end{eqnarray}
Here and in the following, ${\cal O}(1)$ means that it is 
of the order of the characteristic mass scale 
${\rm min}(\Delta m, g\sqrt{c})$. 
The slow-movement 
parameter $\lambda$ in Eq.(\ref{eq:slow-move1}) 
is defined to be of the order of the 
world-volume-coordinate derivative $\partial_\mu$. 
The supertransformation implies that the square of 
the derivative in terms of the Grassmann coordinates $\theta$ 
gives translation in the world-volume : 
$(\partial/\partial\theta)^2 \sim \partial_\mu$. 
Therefore we obtain 
$ d\theta \sim {\partial /\partial \theta }
\sim {\cal O}(\lambda ^{1\over 2})$. 
To assign the order of $\lambda$ for hypermultiplets, 
we observe that the first hypermultiplet $H^1$ has 
nonvanishing values whereas the second hypermultiplet $H^2$ 
vanishes in the $1/2$ BPS background solution (\ref{eq:hyper-sol}). 
If we let the moduli parameters to fluctuate over the 
world-volume coordinates with the order of $\lambda$, 
the fluctuation induces terms of order $\lambda$ in both 
hypermultiplets. 
Therefore the second hypermultiplet $H^2$ naturally 
becomes nonvanishing values and is of order $\lambda$. 
Combining the above order estimates of component fields, 
we assume the order of the hypermultiplet superfields 
and the adjoint chiral scalar superfield 
\begin{eqnarray}
{\bf H}^1\sim {\cal O}(1),\qquad 
{\bf H}^2\sim {\cal O}(\lambda ), \qquad 
{\Phi} \sim {\cal O}(1).
\label{BPS-cond-by-order-assign-wall-1}
\end{eqnarray}
Note that this assignment breaks half of supersymmetry, 
and surviving supersymmetry is manifest in this 
superfield formalism. 
BPS equations for walls also respect 
this supersymmetry automatically, as we will explain later.   
On the other hand, the gauge field $W_\mu$ vanishes in 
the BPS background, and is only induced by the order $\lambda$ 
fluctuations of moduli fields. 
Since the gauge field appears as the coefficient of 
$\bar \theta \gamma^\mu \theta \sim {\cal O}(\lambda^{-1})$, 
we find the vector multiplet to be of the order of 
\begin{eqnarray}
{\bf V}\sim {\cal O}(1),\qquad 
(W_\mu \sim {\cal O}(\lambda )). 
\label{BPS-cond-by-order-assign-wall-3}
\end{eqnarray}
Neglecting ${\cal O}(\lambda ^4)$ 
we obtain 
\begin{eqnarray}
{\cal L}
&=& 
-{\cal E}_{\rm w}
+\int d^4\theta {\rm Tr}\left[
-2c {\bf V}
+e^{2 {\bf V}}{\bf H}^1{\bf H}^{1\dagger }
+{1\over 2g^2}\left(
e^{-2{\bf V}}{\hat D}_ye^{2{\bf V}}\right)^2
\right]\nonumber \\
&&{}+\left(\int d^2 \theta  
{\rm Tr}\left[{\hat D}_y{\bf H}^1{\bf H}^{2\dagger }
-{\bf H}^1M{\bf H}^{2\dagger }\right]+{\rm h.c.}\right). 
\label{eq:Lag-order2}
\end{eqnarray}

Up to this order, we can see that 
${\bf H}^2$, ${\bf V}$ serve as Lagrange multiplier fields. 
Namely the field equations for ${\bf H}^2$ and ${\bf V}$ 
give constraints 
\begin{eqnarray}
\hat D_y{\bf H}^1
&=&{\bf H}^1M, 
\label{eq:superfield-constr2}\\
g^2(c-{\bf H}^1{\bf H}^{1\dagger }e^{2{\bf V}}) 
&=&
- \hat D_y\left(e^{-2{\bf V}}\hat D_ye^{2{\bf V}}\right), 
\label{eq:superfield-constr1}
\end{eqnarray} 
respectively. 
By expanding the superfield constraints 
in powers of the Grassmann coordinates 
$\theta, \bar \theta$, we find, at the leading order, 
the BPS equations  (\ref{BPSeq-H}), and (\ref{BPSeq-Sigma}) 
for the hypermultiplet $H^1$, and vector multiplet 
scalar $\Sigma$ with $H^2=0$. 

We can now choose a convenient gauge of the 
complexified $U(N_{\rm C})$ local gauge invariance. 
Let us define an element of the complexified gauge 
transformation ${\bf S}$ to express the chiral scalar 
superfield ${\Phi}$ for the adjoint scalar of the 
vector multiplet as a pure gauge 
\begin{eqnarray}
{\Phi }={\bf S}^{-1}\partial _y{\bf S}. 
\label{eq:phi-pure-gauge}
\end{eqnarray}
Then the constraint equation 
(\ref{eq:superfield-constr2}) 
for the hypermultiplet chiral 
superfield becomes simpler 
\begin{eqnarray}
 \partial _y({\bf SH}^1)={\bf SH}^1 M ,
\end{eqnarray}
which is easily solved in terms of the moduli 
matrix chiral superfields ${\bf H}_0$ as 
\begin{eqnarray}
{\bf H}^1(x,\theta, \bar \theta,y)
={\bf S}^{-1}(x,\theta, \bar \theta,y)
{\bf H}_0(x,\theta, \bar \theta)e^{My}. 
\label{eq:moduli-matrix-superfield}
\end{eqnarray}

After solving the hypermultiplet constraint equation 
(\ref{eq:superfield-constr2}), 
we can now define a vector 
superfield ${\bfOmega}$ which is 
invariant under the complexified $U(N_{\rm C})$ gauge 
transformations 
\begin{eqnarray}
{\bfOmega }
\equiv {\bf S}e^{-2{\bf V}}{\bf S}^\dagger .
\label{eq:def-Omega}
\end{eqnarray}
The remaining constraint (\ref{eq:superfield-constr1}) 
can be rewritten in terms of the gauge invariant 
superfield ${\bfOmega}$ as 
\begin{eqnarray}
\partial _y\left({\bfOmega }^{-1}
\partial _y{\bfOmega }\right)
=g^2c\left(1-{\bfOmega }^{-1}{\bfOmega }_0\right),
\quad
 {\bfOmega }_0\equiv 
c^{-1}{\bf H}_0e^{2My}{\bf H}_0^\dagger , 
\label{eq:master-eq-superfield} 
\end{eqnarray}
which gives the master equation (\ref{eq:master-eq}) 
as the lowest component. 
Therefore this is the superfield extension of 
the master equation. 

By using the solution of the 
constraint equation (\ref{eq:superfield-constr2}) 
for the hypermultiplet superfield ${\bf H}^1$, 
we can rewrite the fundamental Lagrangian 
in Eq.(\ref{eq:Lag-order2}) 
(up to order ${\cal O}(\lambda^2)$) 
in terms of the gauge invariant 
superfield ${\bfOmega}$ as 
\begin{eqnarray}
{\cal L}&=&
-{\cal E}_{\rm w}
+\int d^4 \theta 
\left[
c\log{\rm det}{\bfOmega} 
+c{\rm Tr}\left({\bfOmega} _0{\bfOmega} ^{-1}\right)
+{1\over 2g^2}
{\rm Tr}\left({\bfOmega }^{-1}
\partial _y{\bfOmega} \right)^2 
\right]
+{\cal O}(\lambda ^4). 
\label{eq:Lag-omega}
\end{eqnarray}
The first, second, and third terms 
in the $d^4 \theta$ integrand 
come from the corresponding terms in the 
$d^4 \theta$ integrand of the fundamental Lagrangian 
(\ref{eq:Lag-order2}) (up to order ${\cal O}(\lambda^2)$).

The superspace extension (\ref{eq:master-eq-superfield}) 
of the master equation  provides a method to determine all 
the quantities of interest as a systematic expansion in powers 
of Grassmann coordinates $\theta, \bar \theta$ as follows. 
Suppose we have an exact solution 
$\Omega _{\rm sol}(H_0,H_0^\dagger ,y)$ 
for the master equation (\ref{eq:master-eq}) 
as a function of moduli matrix $H_0, H_0^\dagger$ 
 \begin{eqnarray}
  \Omega ={\bfOmega }\Big|_{\theta =0}
=\Omega _{\rm sol}(H_0(x),H_0^\dagger (x),y). 
 \end{eqnarray}
By promoting the moduli matrix to a superfield 
${\bf H}_0, {\bf H}^\dagger_0$ 
defined in Eq.(\ref{eq:moduli-matrix-superfield}), 
we obtain the solution for the vector superfield 
${\bfOmega} $ 
of the superfield master equation 
(\ref{eq:master-eq-superfield}) 
as a composite of the chiral and the
anti-chiral superfields, 
\begin{eqnarray}
\Omega _{\rm sol}({\bf H}_0(x,\theta), 
{\bf H}_0^\dagger (x,\bar{\theta}),y)
\equiv 
{\bfOmega }_{\rm sol}
. 
\end{eqnarray}
As we noted in Eq.(\ref{eq:def-Omega}), 
the superfield ${\bfOmega}=
{\bf S}e^{-2{\bf V}}{\bf S}^\dagger$ is $U(N_{\rm C})$ 
supergauge invariant, but the division between ${\bf S}$, 
(${\bf S}^\dagger$) and $e^{-2{\bf V}}$ depends on the 
gauge choice. 
In obtaining the solution for the fluctuation fields such 
as $W_\mu$, we need to choose the Wess-Zumino 
gauge for the real general (vector) superfield 
${\bf V}_{\rm sol}$. 
This gauge transformation to the Wess-Zumino gauge 
is expressed as a multiplication of 
the chiral ${\bf S}_{\rm sol}$ and anti-chiral 
${\bf S}_{\rm sol}^\dagger$ superfields 
from left and right respectively as 
\begin{eqnarray}
{\bf S}_{\rm sol}e^{-2{\bf V}_{\rm sol}}
{\bf S}^\dagger_{\rm sol}
= 
{\bfOmega }_{\rm sol} . 
\label{eq:sol-superfield}
\end{eqnarray} 
Then expansion of the left-hand side of 
(\ref{eq:sol-superfield}) 
in powers of the Grassmann coordinates 
$\theta, \bar \theta$ gives 
\begin{eqnarray}
{\bf S}_{\rm sol}e^{-2{\bf V}_{\rm sol}}
{\bf S}^\dagger_{\rm sol}
= 
S_{\rm sol}S^\dagger _{\rm sol}
+
\theta\sigma^\mu\bar \theta \left(
i(\partial _\mu S_{\rm sol})S_{\rm sol}^\dagger 
-iS_{\rm sol}(\partial _\mu S_{\rm sol}^\dagger )
+2S_{\rm sol}W^{\rm sol}_\mu S_{\rm sol}^\dagger 
\right) +\cdots,
\label{eq:omega-vector}
\end{eqnarray} 
where we have not displayed the bilinear terms of 
fermions, and dots denote other powers of Grassmann 
coordinates. 
Expanding the right-hand side of 
Eq.(\ref{eq:sol-superfield}) we obtain 
\begin{eqnarray}
\Omega _{\rm sol}({\bf H}_0,{\bf H}_0^\dagger ,y)
&=&
\Omega_{\rm sol}
+\theta\sigma^\mu\bar \theta \left(
i(\delta _\mu -\delta _\mu ^\dagger )
\Omega _{\rm sol}\right)
+\cdots , 
\label{eq:omega-sol-vector}
\end{eqnarray}
where we have defined the variation $\delta_\mu$ and its conjugate 
$\delta_\mu^\dagger$ with respect to 
the scalar fields of chiral superfields and 
anti-chiral superfields 
\begin{eqnarray}
\delta_\mu 
\equiv \sum_i\partial_\mu\phi^i
{\delta 
\over\delta \phi^i}, 
\qquad 
\delta_\mu^\dagger 
\equiv \sum_i\partial_\mu \phi^{i*}
{\delta \over \delta \phi^{i*}}, 
\label{eq:variation}
\end{eqnarray}
respectively. 
If the variation $\delta_\mu$ and $\delta^\dagger_\mu$ act 
on those functions which depend on the world-volume 
coordinates $x^\mu$ only through moduli fields, they 
satisfy $\partial _\mu =\delta _\mu +\delta _\mu ^\dagger $.

Comparing the lowest components of (\ref{eq:omega-vector}) 
and (\ref{eq:omega-sol-vector}), we obtain 
\begin{eqnarray}
S_{\rm sol}S^\dagger _{\rm sol}
=
\Omega _{\rm sol}
.
\label{eq:sol-omega}
\end{eqnarray} 
This shows that we cannot avoid 
$S_{\rm sol}$ 
to depend on both $\phi^i$ and $\phi^{i*}$, 
since we cannot factorize these dependences 
in $\Omega_{\rm sol}$. 
One should note that ${\bf S}_{\rm sol}$ 
(${\bf S}_{\rm sol}^\dagger$) is still chiral (anti-chiral) 
scalar superfield, taking both $\phi^i$ and $\phi^{i*}$ 
as lowest components of chiral scalar superfields. 
Comparison of the vector components of (\ref{eq:omega-vector}) 
and (\ref{eq:omega-sol-vector}), we obtain 
a solution of the gauge fields as 
\begin{eqnarray}
- iW_\mu ^{\rm sol}=S^{-1}_{\rm sol}\delta_\mu^\dagger S_{\rm sol}
+S^\dagger _{\rm sol}\delta _\mu S^{\dagger -1}_{\rm sol}
+(\hbox{bi-linear terms of fermions}).
\label{eq:sol-W}
\end{eqnarray}
It is interesting to observe that this solution of gauge 
field fluctuation $W_\mu ^{\rm sol}$ receives contributions 
only from the $\phi^{i*}$ ($\phi^i$) 
dependence of $S_{\rm sol}$ ($S_{\rm sol}^\dagger$), 
in spite of the ${\bf S}_{\rm sol}$ being the chiral 
superfield. 
Similarly the adjoint scalar $\Sigma$ and the gauge 
field $W_y$ in the extra fifth direction is obtained from 
the lowest component of Eq.(\ref{eq:phi-pure-gauge}) 
\begin{eqnarray}
{\Phi }_{\rm sol}
={\bf S}_{\rm sol}^{-1}\partial _y{\bf S}_{\rm sol}
\quad \rightarrow \quad 
\Sigma^{\rm sol} +iW_y^{\rm sol} 
=S^{-1}_{\rm sol}\partial _yS_{\rm sol} .
\end{eqnarray}
The other components of the superfields ${\bfOmega }$, 
${\bf V}$, and ${\Phi }$ 
are similarly determined by the superfield equations.

In order to obtain the low-energy effective Lagrangian 
${\cal L}_{\rm eff}$, we just need to substitute 
the solutions ${\bfOmega}_{\rm sol}$ into the fundamental 
Lagrangian ${\cal L}$ and integrate over the extra 
dimensional coordinate $y$. 
The resulting four-dimensional effective Lagrangian for 
${\bf H}_0$ is given by  
\begin{eqnarray}
{\cal L}_{\rm eff}&=&
\int dy \; {\cal L}
=-T_{\rm w}
+\int d^4\theta K({\bf \phi} ,{\bf \phi} ^*)+{\cal O}(\lambda ^4),
\label{eq:effective-Lag}
\end{eqnarray}
where the K\"ahler potential is expressed by an integral 
form as 
\begin{eqnarray}
 K({\bf \phi} ,{\bf \phi} ^*)&=&\int dy 
{\cal K}({\bf \phi} ,{\bf \phi} ^*,{\bfOmega},y)
\Big|_{{\Omega} ={\Omega} _{\rm sol} },
\label{eq:kahler-pot}
\end{eqnarray}
with a density
\begin{eqnarray}
{\cal K}({\bf \phi} ,{\bf \phi} ^*,{\bfOmega},y)
&=&
c\log{\rm det}{\bfOmega} 
+c{\rm Tr}\left({\bfOmega} _0{\bfOmega} ^{-1}\right)
+{1\over 2g^2}
{\rm Tr}\left({\bfOmega }^{-1}
\partial _y{\bfOmega} \right)^2. 
\label{eq:kahler-density}
\end{eqnarray}
Since we are considering the massless fields corresponding to 
the moduli, we naturally obtain a nonlinear sigma model 
whose kinetic terms are specified by the K\"ahler potential 
$K(\phi, \phi^*)$, without any potential terms. 
Let us note that our method gives the density of the K\"ahler 
potential directly without going through the K\"ahler metric. 
This is in contrast to the component approach where one usually 
obtains the K\"ahler metric of the nonlinear sigma model with 
component scalar fields, 
and then integrate it to obtain 
the K\"ahler potential with a lot of labor. 
It is interesting to note that 
our effective Lagrangian 
is not just an effective Lagrangian on a single wall, but an 
effective Lagrangian on the multiple wall system with 
various moduli such as relative distance moduli as 
the effective fields. 
Therefore we can discuss strings stretched between multiple 
walls (branes), 
which was difficult previously as a BIon \cite{bion}.\footnote{ 
We can construct the full solutions of this composite soliton 
as a 1/4 BPS state \cite{Isozumi:2004vg}.
}

By using the superfield master equation 
(\ref{eq:master-eq-superfield}), 
we can show that 
the second term in Eq.(\ref{eq:kahler-density}) becomes 
a total derivative term. 
Therefore it can be omitted from the 
effective Lagrangian. 
The wall tension $T_{\rm w}$ is given by the topological 
charge in Eq.(\ref{eq:tension}) as an integral over the 
total derivative term ${\cal E}_{\rm w}$ 
in Eq.(\ref{eq:total-der}) 
by using the boundary condition which requires that 
vacua are reached at both infinities $y=\pm \infty$. 

In the strong coupling limit $g^2\rightarrow \infty $, 
the superfield master equation 
(\ref{eq:master-eq-superfield}) 
becomes just an algebraic 
equation $\Omega =\Omega _0$, and 
exact solutions for 
$\Omega $ can be obtained and the K\"ahler potential 
assumes a simple form in this 
case \cite{Isozumi:2004jc} 
\begin{eqnarray}
  K_0(\phi ,\phi ^*)=c\int dy \log{\rm det}\Omega _0. 
\label{eq:kahler-pot-strong}
\end{eqnarray} 

In \cite{Eto:2006uw} we also worked out the integrand of 
the effective K\"ahler potential 
for non-Abelian vortices, 
whose moduli matrix was established in \cite{Eto:2005yh}. 
It turned out to contain the Wess-Zumino-Witten-like term.
On the other hand it has been recently shown that 
the one for domain wall networks (loops) \cite{Eto:2005cp} 
takes the similar form with (\ref{eq:kahler-density}) 
\cite{Eto:2006bb}.
Our method should be applicable to monopoles (instantons).

\end{document}